\documentclass[preprint,12pt]{elsarticle}

\usepackage{amssymb}

 \biboptions{comma,square}

\journal{Physics Letters A}

\begin{document}

\begin{frontmatter}

\title{Zitterbewegung and gravitational Berry phase}

\author{Giorgio Papini}

\address{Department of Physics and Prairie Particle Physics Institute, University of Regina, Regina, Sask S4S 0A2, Canada}
\address{International Institute for Advanced Scientific Studies,
89019 Vietri sul Mare (SA), Italy.}
\ead{papini@uregina.ca}

\begin{abstract}
Berry phases mix states of positive and negative energy in the propagation of fermions and bosons in external gravitational and electromagnetic fields
and generate Zitterbewegung oscillations.
The results are valid in any reference frame and to any order of approximation in the metric deviation.

\end{abstract}


\end{frontmatter}

\section{Introduction}
\label{1}
The contribution of external gravitational fields to the solution of covariant wave equations is contained in a
Berry phase \cite{caipap1}. This should be expected because in general relativity the space of parameters of Berry's theory coincides with space-time.
The wave equations for fermions and bosons  \cite{caipap2}, \cite{caipap3}, \cite{papsc}, \cite{pap1}
have been solved exactly to first order in the metric deviation $\gamma_{\mu\nu}=g_{\mu\nu}- \eta_{\mu\nu}$ for any metric and the solutions
give the correct Einstein deflection when applied to geometrical optics and can be used in interferometry, gyroscopy, in the study of neutrino helicity
and flavour oscillations \cite{pap2} and of spin-gravity coupling \cite{pap3}. They also reproduce a variety of known effects like those
discussed in \cite{COW}, \cite{PW}, \cite{BW},\cite{MASH}.

It is shown below that the gravitational Berry phase gives rise to a field-dependent Zitterbewegung (ZB) in the
propagation of particles in a gravitational background.

\section{Dirac and Klein-Gordon equations}
\label{2}
Consider first the covariant Dirac equation
\begin{equation}\label{CDE}
  [i\gamma^{\mu}(x){\cal D}_\mu-m]\Psi(x)=0\,.
  \end{equation}
The notations are those of \cite{pap2}.
The first
order solutions of (\ref{CDE}) are of the form
\begin{equation}\label{E}
  \Psi(x) = {\hat T}(x) \psi(x)\,,
\end{equation}
where $\psi(x)$ is a solution
of the flat space-time Dirac equation
\begin{equation}\label{DE}
\left(i\gamma^{\hat{\mu}}\partial_{\mu}-m\right)\psi(x)=0\,,
\end{equation}
here a plane wave of four-momentum $k^{\alpha}$ satisfying the relation $k_{\alpha}k^{\alpha}=m^{2}$, and $\gamma^{\hat{\mu}}$ are the usual
constant Dirac matrices. The operator $\hat{T}$ is given by
\cite{pap2}
\begin{equation}\label{T}
    \hat{T}=
  -\frac{1}{2m}\left(-i\gamma^{\mu}(x)\mathcal{D}_{\mu}-m\right)e^{-i\Phi_{T}}\,,
\end{equation}
\begin{equation}\label{PHIS}
\Phi_{T}=\Phi_{S}+\Phi_{G}\,,\qquad
\Phi_{S}(x)=\int_{P}^{x}dz^{\lambda}\Gamma_{\lambda}(z)\,,
\end{equation}
and
 \begin{eqnarray}\label{PH}
  \Phi_{G}(x) = -\frac{1}{4}\int_P^xdz^\lambda\left[\gamma_{\alpha\lambda,
  \beta}(z)-\gamma_{\beta\lambda, \alpha}(z)\right]\left[\left(x^{\alpha}-
  z^{\alpha}\right)k^{\beta}-\left(x^{\beta}-z^{\beta}\right)k^{\alpha}\right]+
\\ \nonumber
 \frac{1}{2}\int_P^xdz^\lambda\gamma_{\alpha\lambda}(z)k^{\alpha}\,,
\end{eqnarray}
where $\Gamma_{\lambda}$ represents the spin connection.
The solutions $\psi(x)$ of (\ref{DE}) can include wave packets, if so desired. In this
case the ZB decays in time \cite{Lock}, which is not an essential point in what follows.
In (\ref{PHIS}) and (\ref{PH}), the path integrals are taken along
the classical world line of the particle starting from a reference point
$P$.

In most applications $\psi(x)$ is represented by a positive energy solution
$\psi(x)=u(\vec{k})e^{-ik_{\mu}x^{\mu}}$. However the influence of negative energy
solutions $\psi^{(1)}(x)=v(\vec{k})e^{ik_{\mu}x^{\mu}}$ can not be neglected
because the wave functions $\psi(x)$ by themselves do not form a complete set.
A relationship
between $\Psi(x)$ and $\Psi^{(1)}(x) = \hat{T}_{1}\psi^{(1)}(x)$ must therefore be found.
The spin-up ($\uparrow$) and spin-down ($\downarrow$) components of the
spinors $u$ and $v$ obey the well-known equations
\begin{equation}\label{spin}
u_{\downarrow}=\gamma^{5}v_{\uparrow}\,,\qquad v_{\downarrow}=\gamma^{5}u_{\uparrow}\,.
\end{equation}
The required relation between $\Psi(x)$ and $\Psi^{(1)}(x)$ follows from (\ref{spin}), or simply from the replacement of $\psi(x)$
with $\gamma^{5}\psi(x)$ in (\ref{DE}).
If, in fact, $\Psi(x)=e^{-ik_{\mu}x^{\mu}}\hat{T}u$ is a solution of (\ref{CDE}), it then follows from (\ref{spin}),
the relations $\left\{\gamma^{5},\gamma^{\hat{\mu}}\right\}
=0$, $\sigma^{{\hat \alpha}{\hat \beta}}=\frac{i}{2}[\gamma^{\hat
\alpha}, \gamma^{\hat \beta}]$,
$  \gamma^\mu(x)=e^\mu_{\hat \alpha}(x) \gamma^{\hat
  \alpha}\,,
  \Gamma_\mu(x)=-\frac{1}{4} \sigma^{{\hat \alpha}{\hat \beta}} e^\nu_{\hat \alpha}e_{\nu\hat{\beta};\, \mu}\,,$
and $    [\gamma^{5},\Gamma^{\mu}]=0$
that
$\Psi^{(1)}(x)=e^{ik_{\mu}x^{\mu}}\hat{T}_{1}v$
also is a solution of (\ref{CDE}) and $\hat{T}_{1}=\gamma^{5}\hat{T}\gamma^{5}$.
It is useful
to further isolate the gravitational contribution in the vierbein
components by writing
$e^{\mu}_{\hat{\alpha}}\simeq\delta^{\mu}_{\hat{\alpha}}+h^{\mu}_{\hat{\alpha}}$, which
leads to
\begin{eqnarray}\label{T}
{\hat
T}=\frac{1}{2m}\left\{\left(1-i\Phi_{G}\right)\left(m+\gamma^{\hat{\alpha}}
k_{\alpha}\right)-i
\left(m+\gamma^{\hat{\alpha}}k_{\alpha}\right)\Phi_{S}+
  \left(k_{\beta}h^{\beta}_{\hat{\alpha}}+
\Phi_{G,\alpha}\right)\gamma^{\hat{\alpha}}\right\}\equiv
\\ \nonumber\hat{T}_{0}+\hat{T}_{G}\,,
\end{eqnarray}
where
$\hat{T}_{0}\equiv\frac{1}{2m}\left(m+\gamma^{\hat{\alpha}}k_{\alpha}\right)$
and $\hat{T}_{G}$ contains the gravitational
corrections. The operator ${\hat T}_1$ can be immediately calculated from (\ref{T}).

The gravitational field mixes the positive and negative energy solutions of
(\ref{DE}). In fact the eigenstates $U^{\pm}=
1/\sqrt{2}(u \pm v)$ of $\gamma^{5}$ and the eigenstates
$u$  and $v$ of $\hat{T}_{0}$ are not the same
and $\hat{T},\,\hat{T}_{1}$ mix $u$  and $v$. The mixing is effected by
$\hat{T}_{G}$ and $\hat{T}_{1G}$ which are entirely due to Berry phase.

 Mixing manifests itself as follows.

The state of a fermion in a gravitational field can be
written in the form
\begin{equation}\label{PS}
 |\Phi(t) \rangle =
  \alpha(t)|\psi(t)\rangle+\beta(t)|\psi^{(1)}(t)\rangle=\alpha_{0}\hat{T}(t)|\psi(t)\rangle+\beta_{0}\hat{T}_{1}(t)|\psi^{(1)}(t)\rangle\,,
\end{equation}
where $|\alpha_{0}|^2+|\beta_{0}|^2=1$,
from which one obtains
\begin{eqnarray}\label{al}
\alpha(t)=
\langle \psi|\Phi(t)\rangle=
\alpha_{0}\langle\psi|\hat{T}|\psi\rangle +\beta_{0}\langle
\psi|\hat{T}_{1}|\psi^{(1)}\rangle\,; \qquad
\\ \nonumber\beta(t)=\langle
\psi^{(1)}|\Phi(t)\rangle =\alpha_{0}\langle
\psi^{(1)}|\hat{T}|\psi \rangle +\beta_{0}\langle
\psi^{(1)}|\hat{T}_{1}|\psi^{(1)}\rangle \,. \qquad
\end{eqnarray}
If at $t=0$ the gravitational field is not present, then
$\hat{T}_{G}=0\,,\hat{T}_{1G}=0$ and
$\alpha(0)\equiv\alpha_{0}=\,,\beta(0)\equiv\beta_{0}$. It follows from
(\ref{al}) that as the system propagates in a
gravitational field, shifts from $|\psi(t)\rangle$ to $|\psi^{(1)}(t)\rangle$
produce oscillations. Thus the geometrical structure of space-time, represented by gravity,
affects Hilbert space by producing oscillations between the positive and negative energy states.

The presence of an electromagnetic field \cite{dinesh} can be accommodated by adding the
term $ qA_{\alpha}$, where $q$ is the charge of the particle,
to $\Phi_{G,\alpha}$ in $\hat{T}$ and
$\hat{T}_{1}$.
The relationship between external electromagnetic fields and ZB has been investigated extensively
by Feschbach and Villars \cite{FV} for both
Dirac and Klein-Gordon equations.

In order to obtain the transition probabilities
$|\alpha(t)|^{2}\,,|\beta(t)|^{2} $ from (\ref{al}) in a concrete
case, one can choose for simplicity
\begin{equation}\label{psi0}
\psi(x)=f_{0,R}e^{-ik_\alpha x^\alpha}=\sqrt{\frac{E+m}{2m}}
  \left(\begin{array}{c}
                f_{R} \\
                 \frac{\sigma^{3} k}{E+m}\, f_{R} \end{array}\right)
                 \,e^{-ik_\alpha x^\alpha}\,,
\end{equation}
where $ f_{R}$ is the positive helicity eigenvector.
The normalizations are $\langle \psi|\psi\rangle =1$, where $\langle
\psi|= \langle\psi^{\dag}|\gamma^{\hat{0}}$, $\langle
\psi^{(1)}|\psi^{(1)}\rangle =-1$ and $\langle
\psi|\psi^{(1)}\rangle = \langle \psi^{(1)}|\psi\rangle =0$. In
addition, one needs explicit expressions of the metric components
for the purpose of calculating $\hat{T}$ and $\hat{T}_{1}$. The
choice of the metric
\begin{equation}\label{LTmetric}
 \gamma_{00}=2\phi\,, \quad \gamma_{ij}=2\phi\delta_{ij}\,, \quad
 \end{equation}
where $\phi=-\frac{GM}{r}$,
and \textit{M}, \textit{R},
are mass and radius of the source, is again dictated by simplicity.
The vierbein components to order
$\mathcal{O}(\gamma_{\mu\nu})$ are given by
\begin{equation}\label{3.5}
 e^0_{\hat{i}}=0\,{,}\quad
 e^0_{\hat{0}}=1-\phi\,{,}\quad
 e^i_{\hat{0}}=0 \,{,}\quad
 e^l_{\hat{k}}=\left(1+\phi\right)\delta^l_k\,.
 \end{equation}
 Without loss of generality, one may consider particles starting from
 $z=-\infty$, and propagating
 along $x=b\geq R\,,y=0$ in the field of the gravitational source and set $k^{3}\equiv k$ and $k^{0}\equiv E$.

Returning to (\ref{al}), if originally the system is in a
positive energy state, then $ \alpha_{0}=1\,, \beta_{0}=0$, $
|\Phi(t)\rangle = \hat{T} |\psi\rangle $ and from
(\ref{PS}) and $\Phi_{G,3}=(E^{2}/k +k)\phi$ one gets
\begin{equation}\label{al2}
\beta(t)=
\frac{e^{-2iq_{\alpha}x^{\alpha}}}{2m}\left\{-\langle \psi^{(1)}|
\left[Eh_{\hat{0}}^{0}\gamma^{\hat{0}}+\left(-kh_{\hat{3}}^{3}+\left(\frac{E^2}{k}+k\right)
\phi(z)\right)\gamma^{\hat{3}}\right]|\psi\rangle\right\}\,,
\end{equation}
where $q_{0}\equiv E$ because the field does not depend on time, hence energy is conserved, and $q_{i}\equiv k_{i}^{(i)}-k_{i}^{(f)}$.
The first two terms in (\ref{al2}) are due to $\Gamma_{\mu}$ and refer to $\Phi_{S}$. The remaining two terms
come from $\Phi_{G,3}$ and are also Berry phase contributions.
Thus, according to (\ref{al}) and (\ref{al2}), the propagation of the particle has two overlapping components:
one in which the state of the particle does not change, the other
in which oscillations take place from and to energy states of opposite sign with a frequency $2E$, or in ordinary units $2E/\hbar$.
This is at least as large as the ZB frequency $2m$. The particle therefore behaves
as if it were trying to conserve energy-momentum and angular momentum during its propagation.
The presence of the gravitational Berry phase translates into a ZB
that vanishes when there is no gravity acting on the particle and is therefore due to a real force,
as pointed out in \cite{FV} for the case of an external electromagnetic field.
Because the approach is covariant, the result holds true in any frame of reference.
Moreover, the non-local potential $K_{\lambda}(x,x_{0})=\Phi_{G,\lambda}(x, x_{0})+\Gamma_{\lambda}(x)$ can be calculated to any order,
meaning that a ZB also exists at any order.

The transition amplitude $\langle\psi|\hat{T}|\psi\rangle$ can be better calculated using the relation $\langle\psi|\hat{T}|\psi\rangle=\int_{\lambda_{0}}^{\lambda}\langle\psi|\dot{x}^{\mu}\partial_{\mu}\hat{T}|\psi\rangle d\lambda$,
where $\dot{x}^{\mu}=k^{\mu}/m$ and $\lambda$ is an affine parameter along the particle world line. The calculation is outlined in \cite{pap2}.

The probability of the transition $\psi \rightarrow \psi^{(1)}$
follows from (\ref{al2}) and is
\begin{equation}\label{BL}
P_{\psi\rightarrow \psi^{(1)}}=|\beta(t)|^{2} =\left[\frac{1}{2m^2} (k^2 - \frac{E^3}{k})\right]^2 \phi^2(z) \,.
\end{equation}

If $ \alpha_{0}=0, \beta_{0}=1$, then $|\alpha(t)|^2$
represents the probability for the inverse process
$\psi^{(1)}\rightarrow \psi$. One finds
\begin{eqnarray} \label{BE}
P_{\psi^{(1)}\rightarrow \psi}=|\alpha(t)|^2=|\langle \psi|\hat{T}|\psi^{(1)}\rangle|^{2}=
|\langle \psi|\gamma^{5}\hat{T_{1}}\gamma^{5}|\psi^{(1)}\rangle|^{2}= \qquad
\\ \nonumber|\langle \psi^{(1)}|\hat{T}_{1}|\psi\rangle|^{2}=P_{\psi\rightarrow \psi^{(1)}}\,.
\end{eqnarray}
According to (\ref{BL}) and (\ref{BE}), the transitions proceed in both directions with
the same probability, as expected.

As mentioned above, an external electromagnetic field can be introduced by simply adding the corresponding Berry phase to (\ref{al2}).
The additional term in curly brackets is therefore $\langle \psi^{(1)}|-q A_{\mu}\hat{\gamma}^{\mu}|\psi\rangle$. If the addition corresponds to an
electromagnetic wave of amplitude $f$ and frequency $\omega$, in vanishing gravity, there is a
resonance at $\omega =2E$ that leads to $|\beta(t)|^2=(qkf/m^2 \omega)^2 \cos^2(2Ex_{0})$.
If gravity is also present, the resonance condition becomes $\omega=2E$, $C=qkf/m^2 \omega\equiv A$, with $C$ represented by the terms of (\ref{al2}) in curly
brackets, and $|\beta(t)|^2=(A/m)^2 \sin^2(Et)$. The prospects of achieving resonance in laboratory conditions in the near future do not appear favourable.

ZB appears to be universal in condensed matter physics and is the subject of recent, intense research \cite{ZR}. It is in this area that lie
the best opportunities to observe ZB.

Entirely similar conclusions can be reached for the covariant Klein-Gordon equation
\begin{equation}\label{CKG}
\left(g^{\mu\nu}\nabla_{\mu}\nabla_{\nu}+m^2\right)\Phi(x)=0\equiv \hat{\mathcal{T}}\Phi(x)\,,
\end{equation}
which has the first order solution
\begin{equation} \label{SCKG}
\Phi(x) = e^{-i\Phi_{G}}\varphi(x)\,,
\end{equation}
where $\varphi(x)$ satisfies the Klein-Gordon equation in flat space-time
\begin{equation}\label{KG}
\left(\eta^{\mu\nu}\partial_{\mu}\partial_{\nu}+m^2\right) \varphi(x)=0\equiv\hat{\mathcal{T}}_{0}\varphi(x)\,.
\end{equation}
Following the procedure of \cite{FV} one can write the plane wave solutions of (\ref{KG}) as
\begin{equation}\label{FV}
\varphi^{+}(x)=e^{-ip_{\mu}x^{\mu}}\chi^{+}(p)\,\,,\,\,
\varphi^{-}(x)=e^{ip_{\mu}x^{\mu}}\chi^{-}(p)\,,
\end{equation}
where $\chi^{\pm}(p)$ are known functions of $p$.
Representing a generic state of the system $\Lambda(x)$ in terms of the free-field
solutions $\varphi^{\pm}$ one sees immediately that $\Lambda(x)$ is not an eigenstate of $\hat{\mathcal{T}}_{0}$ and that,
therefore, the gravitational part $\hat{\mathcal{T}}-\hat{\mathcal{T}}_{0}$ due to $\Phi_{G}$
mixes the states of positive and negative energy.

Similar results can be obtained for all known relativistic wave equations.

\section{Summary and discussion}
\label{3}
It was shown in \cite{FV} that static electric and magnetic fields in flat space-time excite a field-dependent ZB.
This result has been extended, in this work, not only to electromagnetic fields of any type in curved space-time, but
also to any gravitational fields of weak to intermediate strength. The extension is based on the notion of
Berry phase. Since the approach is covariant, the result holds true in any reference
frame. Moreover, the gauge potential $K_{\lambda}(x, x_{0})$ exists to any order, hence the results remain valid to any order
of approximation in $\gamma_{\mu\nu}$ for both fermions and bosons.

Particle propagation is affected by gravitational and electromagnetic Berry phases. They imply gauge structures that mix the field-free states giving rise
to oscillations of frequency at least as high as $2m$.
This action can be interpreted, in the gravitational case, as an example of how the curvature of space-time can affect Hilbert
space by determining transitions between states of positive and negative energy. The transitions involve $\hbar$.
Though resonance conditions between ZB and the external fields exist in principle, their realization for particles in vacuum
seems unlikely at present. The significance of the results is related to the role played by Berry phase and the related potential $K_{\lambda}(x, x_{0})$
in the mixing of positive and negative energy states that are necessarily contained in the eigenfunctions of relativistic
particles. ZB oscillations appear as the particles strive to conserve energy-momentum and angular momentum
along their world lines.

\bibliographystyle{elsarticle-num}
\bibliography{<your-bib-database>}

\end{document}